\documentclass{article}

\usepackage{arxiv}
\usepackage[utf8]{inputenc}
\usepackage[T1]{fontenc}
\usepackage{hyperref}
\usepackage{url}
\usepackage{booktabs}
\usepackage{amsfonts, amsmath}
\usepackage{nicefrac}
\usepackage{microtype}
\usepackage{graphicx}
\usepackage{natbib}
\usepackage{doi}
\usepackage{enumitem}

\usepackage{todonotes}
\usepackage[nolist]{acronym}

\usepackage{xr-hyper}
\usepackage{xcolor}        

\makeatletter
\newcommand*{\addFileDependency}[1]{%
	\typeout{(#1)}\@addtofilelist{#1}%
	\IfFileExists{#1}{}{\typeout{No file #1.}}}
\newcommand*{\myexternaldocument}[1]{%
	\externaldocument{#1}%
	\addFileDependency{#1.tex}%
	\addFileDependency{#1.aux}}
\makeatother
\myexternaldocument{PINNsupplement}

\setlist[itemize]{leftmargin=*}

\begin{acronym}
	\acro{PINN}{Physics-Informed Neural Network}
	\acro{gLV}{generalised Lotka–Volterra}
	\acro{RMSE}{Root-Mean-Squared Error}
	\acro{MC}{Monte-Carlo}
	\acro{UQ}{Uncertainty Quantification}
	\acro{IC}{initial condition}
\end{acronym}

\title{Exploring Accuracy and Uncertainty Quantification in Physics-Informed Neural Networks for Inferring Microbial Community Dynamics}

\author{%
	\href{https://orcid.org/0000-0002-0535-2684}{\includegraphics[scale=0.06]{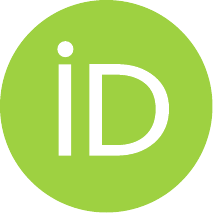}\hspace{1mm}Pedro Fontanarrosa}\thanks{Correspondence: \texttt{pfontanarrosa@gmail.com}} \\
	Cell and Developmental Biology Department, University College London, UK
	\And
	\href{https://orcid.org/0000-0002-9459-1395}{\includegraphics[scale=0.06]{orcid.pdf}\hspace{1mm}Chris P.~Barnes} \\
	Cell and Developmental Biology Department, University College London, UK
}

\hypersetup{
	pdftitle={Exploring Physics-Informed Neural Networks for Microbial Dynamics},
	pdfsubject={q-bio.QM},
	pdfauthor={Pedro Fontanarrosa, Chris P. Barnes},
	pdfkeywords={Physics-informed neural networks, Microbiome, gLV, Uncertainty quantification, Ensemble learning},
}

\graphicspath{{.}{figures/}{fig/}}

\begin{document}
	\maketitle
	
	\begin{abstract}
		\acp{PINN} have become a popular way to infer interpretable interaction parameters from noisy microbial time series, but practitioners face many tunable design choices (loss weights, regularisers, scaling, training schedules) with little guidance, and uncertainty is rarely quantified. We present a two-part study using a DeepXDE \ac{PINN} for a six-species \ac{gLV} model with a known step input $u(t){=}\mathbf{1}_{t\ge5}$ and fixed, species-specific step amplitudes~$\boldsymbol\varepsilon$, under synthetic noise $\sigma{=}0.30$. 
		
		\textbf{Part A—Accuracy ablation.} Starting from a broad-init baseline (RMSE: $\mu{=}0.405$, $M{=}1.152$, $\varepsilon{=}0.456$), we systematically vary constrained initialisation, parameter scaling, L2 regularisation with loss-weighting, split training with auxiliary observations, function constraints, adaptive collocation, hyperparameter tuning, and optimisers (Adam$\rightarrow$L\mbox{-}BFGS). The largest gains come from simple parameter scaling ($M{\times}10$, $\varepsilon{\times}5$ during training, later unscaled), reducing RMSE to $\mu{=}0.196$ ($-51.7\%$), $M{=}0.021$ ($-98.2\%$), and $\varepsilon{=}0.025$ ($-94.5\%$). Hyperparameter tuning and adaptive collocation achieve competitive $M$ errors ($0.025$–$0.027$).
		
		\textbf{Part B—Uncertainty quantification.} We benchmark deep ensembles and Monte Carlo (MC) dropout and report multi-replicate fits with different initial conditions. An $N{=}10$ deep ensemble yields the best single-trajectory growth-rate accuracy ($\mu{=}0.185$, $-54.2\%$) with $M{=}0.024$ and $\varepsilon{=}0.045$, while MC-dropout provides predictive bands. Overall, scaling plus either tuning or ensembles delivers robust interaction inference (near-$2\times10^{-2}$ RMSE for $M$); our loss-weight sweep and additional regularisation settings offer practical defaults. We release code, metrics, and figures to serve as baselines for accuracy and \ac{UQ} in microbiome \acp{PINN}. 
	\end{abstract}

	
	\section{Introduction}\label{sec:intro}
	\acresetall
	\acp{PINN} approximate solutions to differential equations by embedding physical laws into a neural network loss function \cite{raissi2019physics}. Combining mechanistic and data-driven models, they have shown promise in biological modelling \citep{yazdani_2020_SystemsBiologyInformed,daneker_2023_SystemsBiologyIdentifiability}. Applications to microbial interaction inference remain relatively sparse~\cite{bucci2016mdsine,venturelli2018deciphering,clark2021design}. Here we evaluate practical training choices and \ac{UQ} add-ons for \ac{gLV}-based \acp{PINN} in the noisy, single-trajectory regime that typifies microbiome time series, and we extend to multi-replicate fits and predictive uncertainty.
	
\subsection{Contributions}
\begin{itemize}
    \item A reproducible \ac{PINN} baseline that jointly infers growth rates $(\boldsymbol\mu)$, the interaction matrix $(\mathbf M)$, and \emph{species-specific step amplitudes} $(\boldsymbol\varepsilon)$ under a known exogenous input $u(t){=}\mathbf{1}_{t\ge5}$.
    \item An eight-technique single-trajectory ablation with real numbers: constrained initialisation, parameter scaling, L2 + loss-weighting, split training with auxiliary observations, function constraints, adaptive collocation, hyperparameter tuning, and deep ensembles.
    \item New results: $\lambda$-sweep for physics weighting, Adam$\rightarrow$L-BFGS training, L2 penalty on parameters, \emph{multi-replicate} inference under different initial conditions, and uncertainty via deep ensembles and MC-dropout.
\end{itemize}
	
\section{Materials and Methods}\label{sec:methods}
    \subsection{Generalised Lotka-Volterra model}\label{sec:glv}

    We use a \ac{gLV} system with a known exogenous step input \(u(t)=\mathbf{1}_{\,t\ge t_0}\) (with \(t_0=5\)) and a fixed, species-specific perturbation amplitude vector \(\boldsymbol\varepsilon\in\mathbb{R}^S\):
    \begin{align}
    \frac{d x_i}{dt}
      &= x_i \left( \mu_i + \sum_{j=1}^S M_{ij}\, x_j + \varepsilon_i\,u(t) \right),
      \qquad i=1,\dots,S,\\[4pt]
    \dot{\mathbf{x}}(t)
      &= \mathbf{x}(t) \odot \Big( \boldsymbol{\mu} + \mathbf{M}\,\mathbf{x}(t) + \boldsymbol{\varepsilon}\,u(t) \Big), 
    \qquad u(t)=\mathbf{1}_{\,t\ge 5}.
    \end{align}
    Here \(x_i\) is the abundance of species \(i\), \(\boldsymbol{\mu}\) the intrinsic growth rates, \(\mathbf{M}\) the interaction matrix, and \(\boldsymbol{\varepsilon}\) a \emph{constant} vector of per-species step amplitudes. We infer \(\theta=\{\boldsymbol{\mu},\mathbf{M},\boldsymbol{\varepsilon}\}\) from data; the input \(u(t)\) is known and fixed.

    \subsection{Synthetic Data Generation}\label{sec:data}
        We simulate six-species \ac{gLV} systems on $t\!\in\![0,10]$ with a hard step perturbation
        $u(t){=}\mathbf{1}_{t\ge5}$ and sample 101 uniformly spaced time points. Ground-truth parameters are drawn i.i.d.\ as:
        \[
        \mu_i \sim \mathcal U(0.8,\,1.6),\quad
        M_{ii} \sim \mathcal U(-0.16,\,-0.04),\quad
        M_{ij\neq i} \sim \mathcal U(-0.03,\,0.06),\quad
        \varepsilon_i \sim \mathcal U(-0.25,\,0.25).
        \]
        Trajectories are integrated with \texttt{scipy.integrate.odeint}; any realisation with NaN/Inf or
        $\max_t \max_i x_i(t) \ge 10^3$ is rejected and resampled. Observations are corrupted with additive Gaussian noise
        $x_i^{\text{obs}}(t)=x_i(t)+\eta_i(t)$, $\eta_i(t)\!\sim\!\mathcal N(0,\,\sigma^2)$ with $\sigma{=}0.30$.
        
        We use three generators corresponding to our experiments:
        \begin{itemize}
          \item \textbf{Single-trajectory (Study A).} For each of the 50 simulations, we fix the \ac{IC} to
          $x_i(0)=10$ for all species (vector of tens) and draw $(\boldsymbol\mu,\mathbf M,\boldsymbol\varepsilon)$ once; we then add noise to obtain one observed trajectory.
          \item \textbf{Same-IC replicates.} For 50 simulations, we again use $x_i(0)=10$ and draw $(\boldsymbol\mu,\mathbf M,\boldsymbol\varepsilon)$ once; we create $3$ noisy replicates by adding independent noise to the same clean solution.
          \item \textbf{Different-IC replicates (multi-replicate).} For 50 simulations, we draw $(\boldsymbol\mu,\mathbf M,\boldsymbol\varepsilon)$ once and generate $3$ replicates with distinct initial conditions
          $x_i(0)\sim \mathcal U(5,\,15)$ independently per replicate; each replicate then receives independent observation noise.
        \end{itemize}
        The above corresponds to the released scripts \texttt{simulations.py} (single), \texttt{sim\_sameIC.py} (same-IC), and \texttt{sim\_diffIC.py} (different-IC).

	
	\subsection{Baseline PINN Architecture}\label{sec:baseline}
	\begin{itemize}
		\item \textbf{Framework:} DeepXDE with TensorFlow (\texttt{tf.compat.v1} graph mode).
		\item \textbf{Network:} 3 hidden layers, 128 neurons each, \texttt{swish} activations; Glorot normal initialisation.
		\item \textbf{Loss Function:} We use a standard PINN objective decomposed into a data term, a physics (gLV) residual, and optional auxiliary/BC terms \citep{raissi2019physics,yazdani_2020_SystemsBiologyInformed,lu2021deepxde}:
\begin{align}
\mathcal L
&= \underbrace{\frac{1}{N}\sum_{n=1}^N \bigl\|\hat{\mathbf x}(t_n)-\mathbf y_n\bigr\|_2^2}_{\mathcal L_{\text{data}}}
\;+\;
\lambda_{\text{pde}}\,
\underbrace{\frac{1}{N_c}\sum_{m=1}^{N_c}
\bigl\|\mathbf r(\tau_m)\bigr\|_2^2}_{\mathcal L_{\text{pde}}}
\;+\;
\lambda_{\text{aux}}\,
\underbrace{\frac{1}{N_a}\sum_{k=1}^{N_a}
\bigl\| \mathbf A_k \hat{\mathbf x}(\tilde t_k)-\mathbf b_k \bigr\|_2^2}_{\mathcal L_{\text{aux}}}, \label{eq:pinloss}
\end{align}
with residual
\begin{align}
\mathbf r(t)
&= \dot{\hat{\mathbf x}}(t)
- \hat{\mathbf x}(t)\odot\Bigl(\boldsymbol\mu + \mathbf M\,\hat{\mathbf x}(t) + \boldsymbol\varepsilon\,u(t)\Bigr),
\qquad u(t)=\mathbf 1_{\,t\ge 5}.
\end{align}
Here $\dot{\hat{\mathbf x}}(t)$ is obtained by automatic differentiation and $\odot$ is the elementwise product.
$\mathcal L_{\text{data}}$ fits observations, $\mathcal L_{\text{pde}}$ enforces the gLV dynamics at collocation points, and $\mathcal L_{\text{aux}}$ (used in split/auxiliary training) encodes simple anchors such as $t{=}0$ or mid/final time constraints.
(When stated, we also add $\lambda_\theta\|\theta\|_2^2$; $M$ and $\varepsilon$ are internally rescaled during optimisation and unscaled for reporting).

		\item \textbf{Training:} Adam (20k iterations; learning rate per experiment), optionally followed by L-BFGS.
		\item \textbf{Perturbation:} Known step input $u(t)$; $\boldsymbol\varepsilon$ (species-specific step amplitudes) learnt jointly with $\boldsymbol\mu$ and $\mathbf M$.
	\end{itemize}
	
	\subsection{Parameter Scaling}\label{sec:scaling}
	For better-conditioned gradients, we scale during optimisation as $M\!\leftarrow\!M\cdot s_M$ and $\varepsilon\!\leftarrow\!\varepsilon\cdot s_\varepsilon$ with $(s_M,s_\varepsilon){=}(10,5)$, and unscale for reporting.
	
	\subsection{Loss-Weighting and Regularisation}\label{sec:loss_reg}
	We explore: (i) L2 weight decay on NN layers; (ii) a PDE-residual weight $\lambda_{\text{pde}}$; and (iii) an L2 penalty directly on $(\mu,M,\varepsilon)$ (unscaled) in one variant.
	
	\subsection{Training Strategies}\label{sec:training_strategies}
	\paragraph{Split training + auxiliary observations.} We use a two-stage schedule. \emph{Stage 1} (supervised-only) runs 1{,}000 Adam iterations (\(\approx 5\%\) of the 20{,}000 total) with the PDE residual switched off; the loss is driven by (i) all observation times as PointSetBCs and (ii) two simple auxiliary anchors per species at \(t{=}0\) and at the series mid-point \(t{=}5\). \emph{Stage 2} re-enables the physics term (\(\lambda_{\text{pde}}{=}1\)) and continues for the remaining 19{,}000 iterations, keeping the same data and auxiliary terms active.
	\paragraph{Adaptive collocation.} After a warm-up of 10{,}000 Adam iterations on the original 101 anchors (the observation times), we evaluate the PDE residual on a dense grid of 1{,}000 candidate time points over \([0,10]\), select the 200 with the largest residual norm, and add them as extra anchors (total anchors \(=\) 101\(+\)200). Training then continues for a further 10{,}000 iterations with the augmented anchor set.
	\paragraph{Function constraints.}
	Reparameterisations enforce $\mu_i \ge 0$ and $M_{ii} \le 0$ via softplus transforms
	($\mu_i=\mathrm{softplus}(\phi_{\mu,i})$, $M_{ii}=-\mathrm{softplus}(\phi_{M,ii})$),
	reflecting biological priors of non-negative intrinsic growth and negative self-interaction
	(self-limitation/density dependence). Off-diagonal interactions $M_{ij}$ are left
	unconstrained unless stated.
	
	\subsection{Multiple Replicates \& Uncertainty}\label{sec:uq_methods}
	\paragraph{Multi-replicate aggregation.} For simulations with multiple noisy replicates (different initial conditions), we fit one PINN per replicate and average inferred parameters across replicates.
	\paragraph{Deep ensembles.} $N{=}10$ independent initialisations; parameters aggregated as mean $\pm$ SD; trajectories reported with mean $\pm$ SD bands.

\paragraph{MC-dropout.} MC-dropout is a method where dropout is kept active at inference time and multiple stochastic forward passes are performed, treating the variability in predictions as an approximation of Bayesian uncertainty \citep{gal2016_dropout}. This approach has also previously been used in the context of \acp{PINN} \citep{zhang2019_total_uncertainty_pinns}. We performed 100 stochastic passes yielding predictive bands (mean$\pm$SD).

	\section{Results}\label{sec:results}
	
	\subsection{Baseline and Simple Enhancements}
	With broad initialisation and no scaling/constraints (Baseline), average RMSE across simulations was $\mu{=}0.405$, $M{=}1.152$, and $\varepsilon{=}0.456$. Constrained initialisation drastically improved $M$ and $\varepsilon$ (to $0.072$ and $0.067$) but only slightly helped $\mu$ ($0.392$).
	
	\textbf{Parameter scaling} (training with $M\times10$, $\varepsilon\times5$) delivered the largest overall gain: $\mu{=}0.196$, $M{=}0.021$, and $\varepsilon{=}0.025$. \textbf{L2 + $\lambda_{\text{pde}}$} and \textbf{tuning} were close for $M$ ($\approx 0.025$) and improved $\varepsilon$ to $0.047$–$0.067$, with $\mu{=}0.214$–$0.233$.
	
	\subsection{Training Strategies}\label{sec:train_strat_results}
	\textbf{Adaptive collocation} achieved $M{=}0.027$ with $\mu{=}0.227$ and $\varepsilon{=}0.061$. \textbf{Split training + auxiliary observations} and \textbf{function constraints} remained better than baseline for $M$ but were less competitive for $\mu$ and $\varepsilon$ in these settings.
	
	\subsection{Ensembles and Robustness (single trajectory)}\label{sec:ens_results}
	A \textbf{deep ensemble} ($N{=}10$ seeds) gave the best growth-rate accuracy ($\mu{=}0.185$) while keeping $M$ and $\varepsilon$ low ($0.024$ and $0.045$). Ensembles stabilised across seeds without changing the model class.
	
	\subsection{Ablation Summary (single trajectory)}\label{sec:ablation}
\begin{figure}[p]
		\centering
		\includegraphics[width=0.8\linewidth]{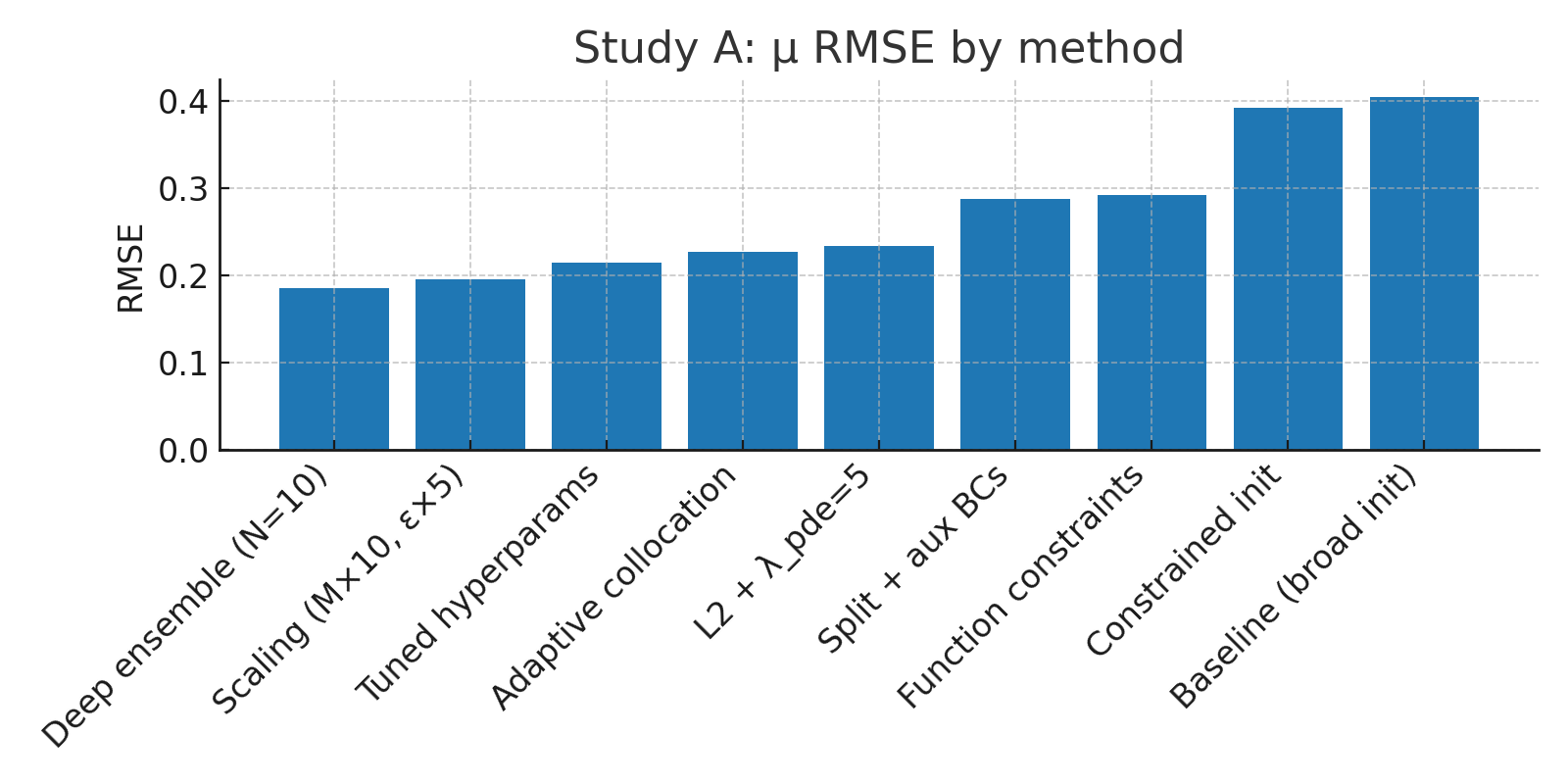}\\
		\includegraphics[width=0.8\linewidth]{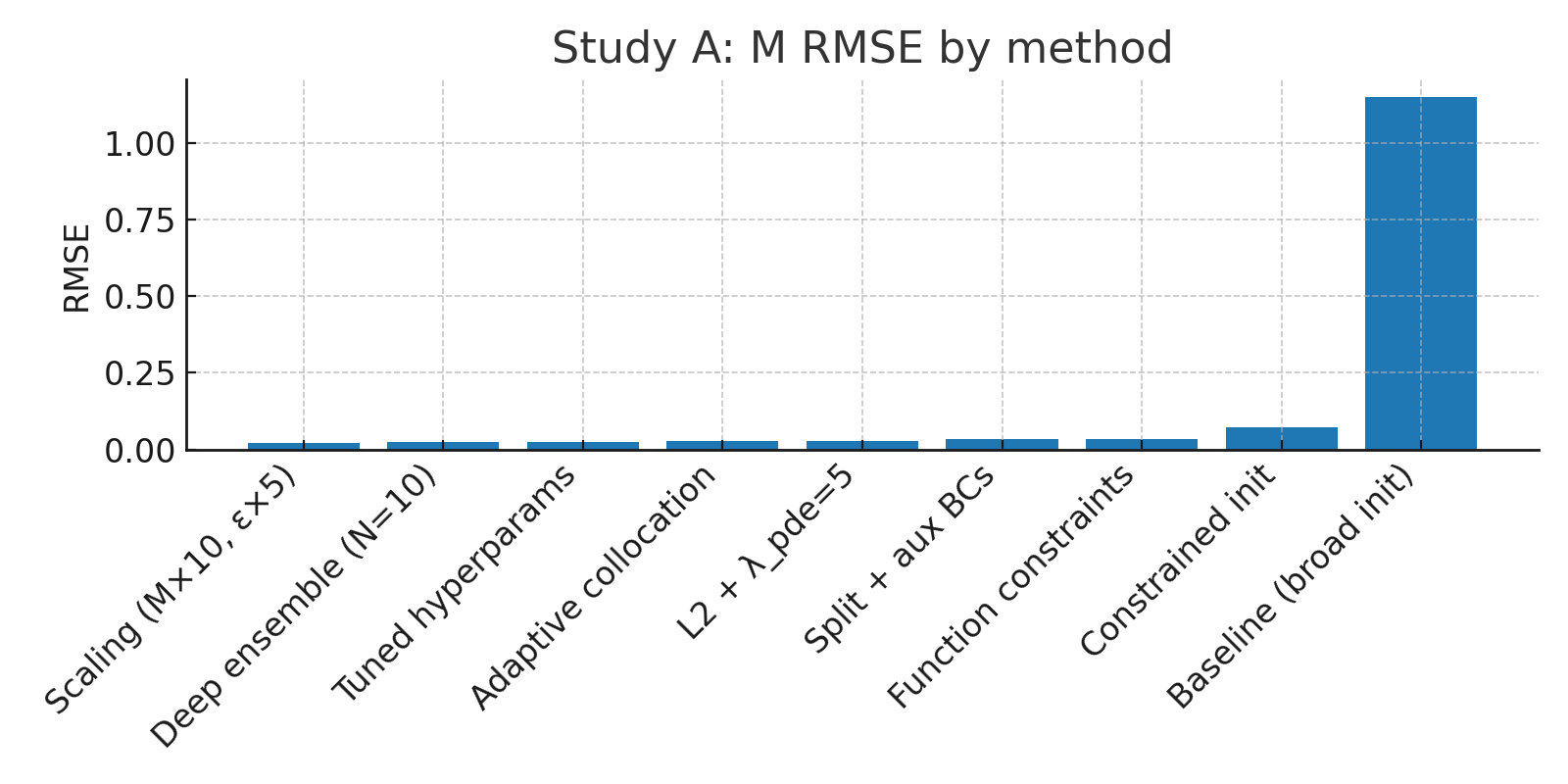}\\
		\includegraphics[width=0.8\linewidth]{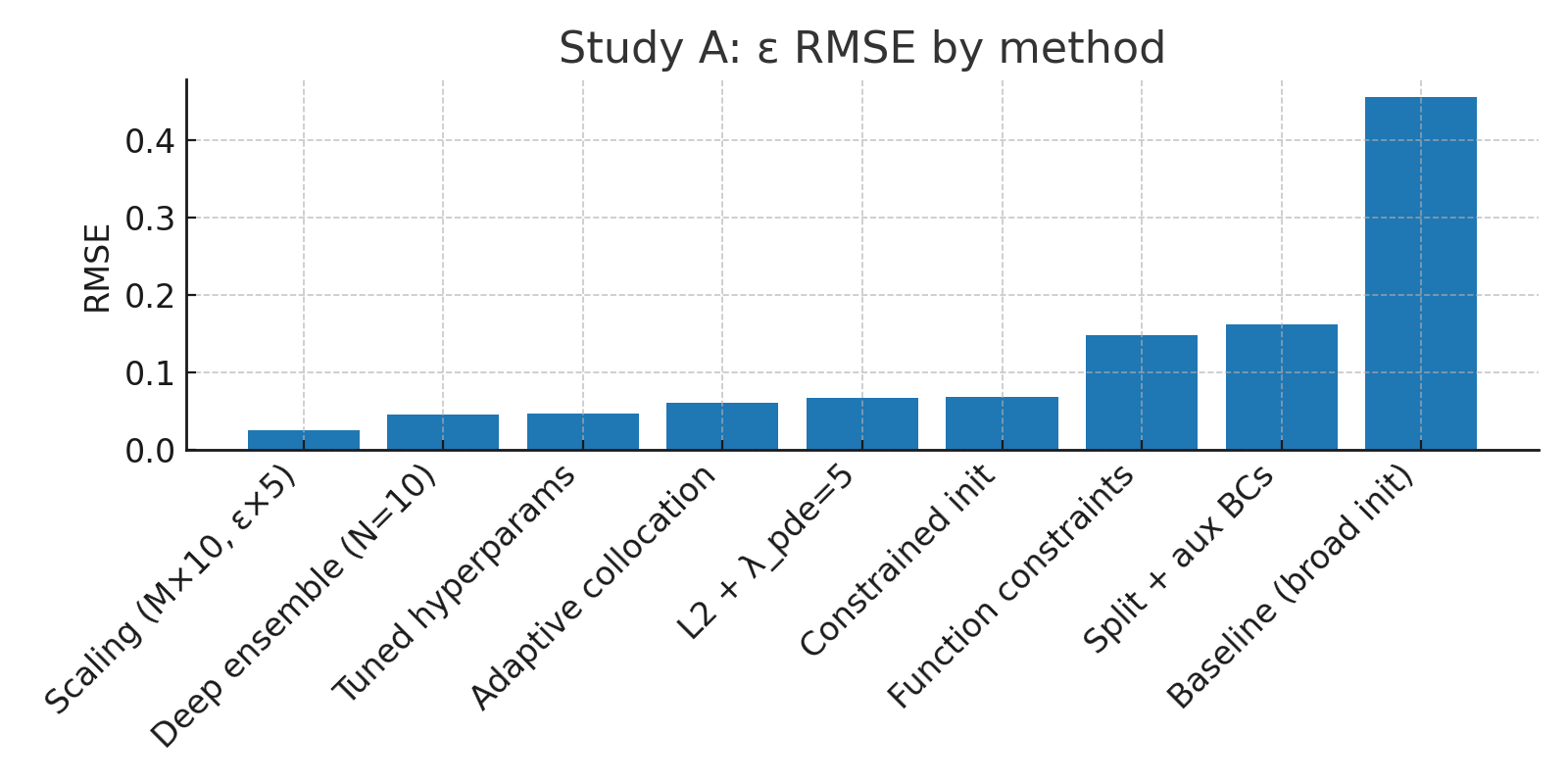}
		\caption{Study A (single trajectory): RMSE by method for $\mu$, $M$, and $\varepsilon$.}
		\label{fig:studyA_rmse}
	\end{figure}

\begin{figure}[p]
		\centering
		\includegraphics[width=0.8\linewidth]{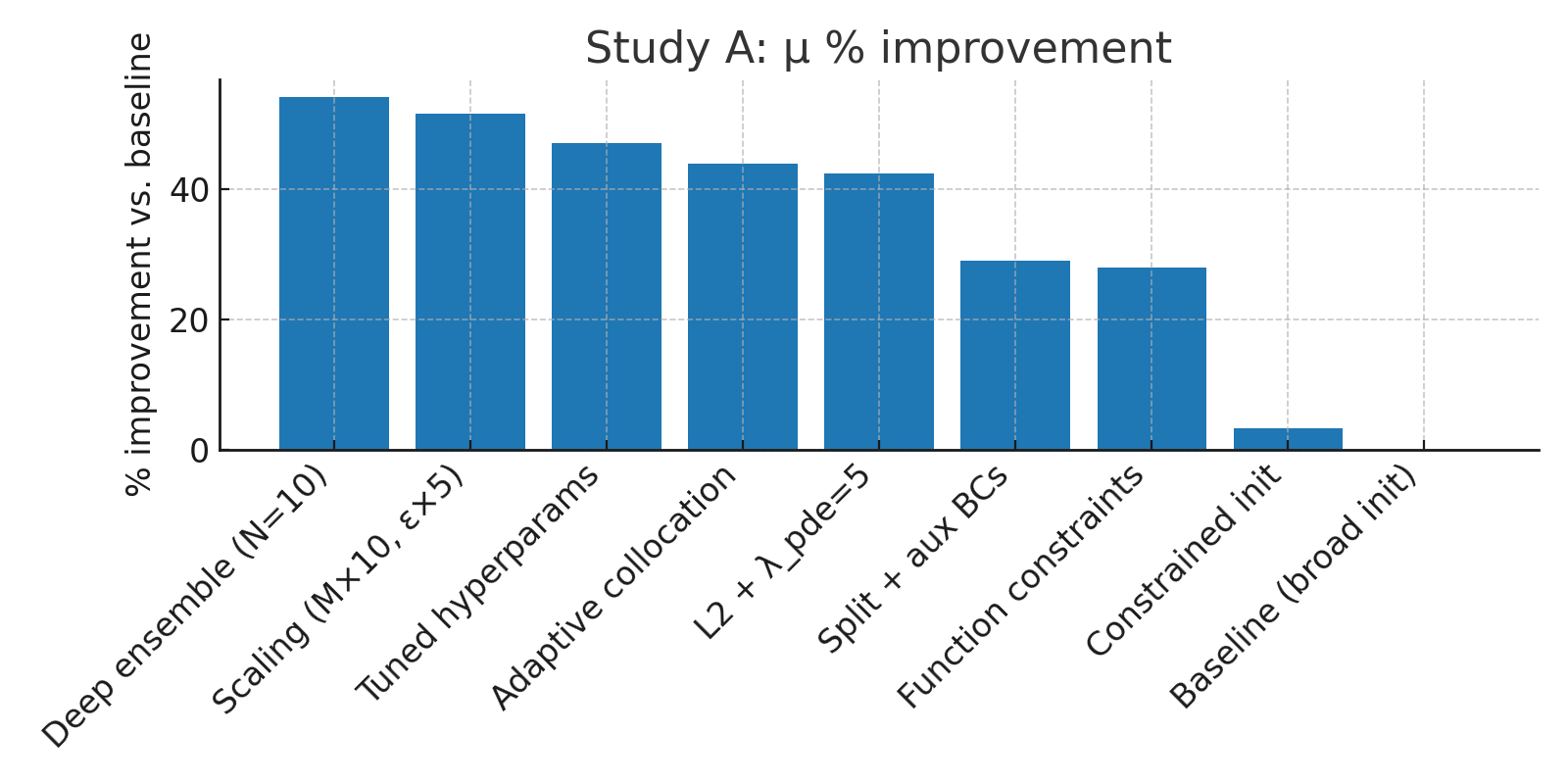}\\
		\includegraphics[width=0.8\linewidth]{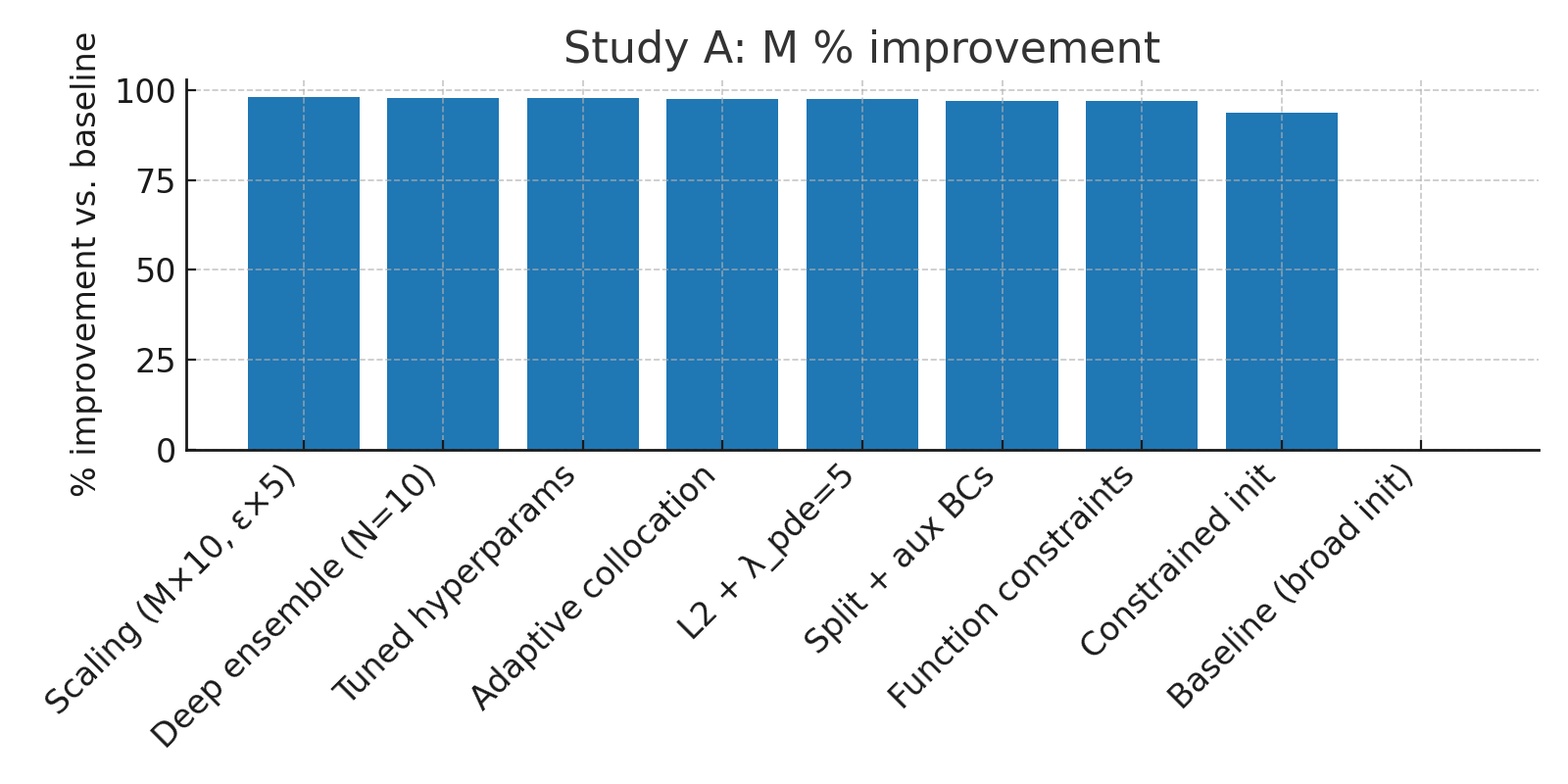}\\
		\includegraphics[width=0.8\linewidth]{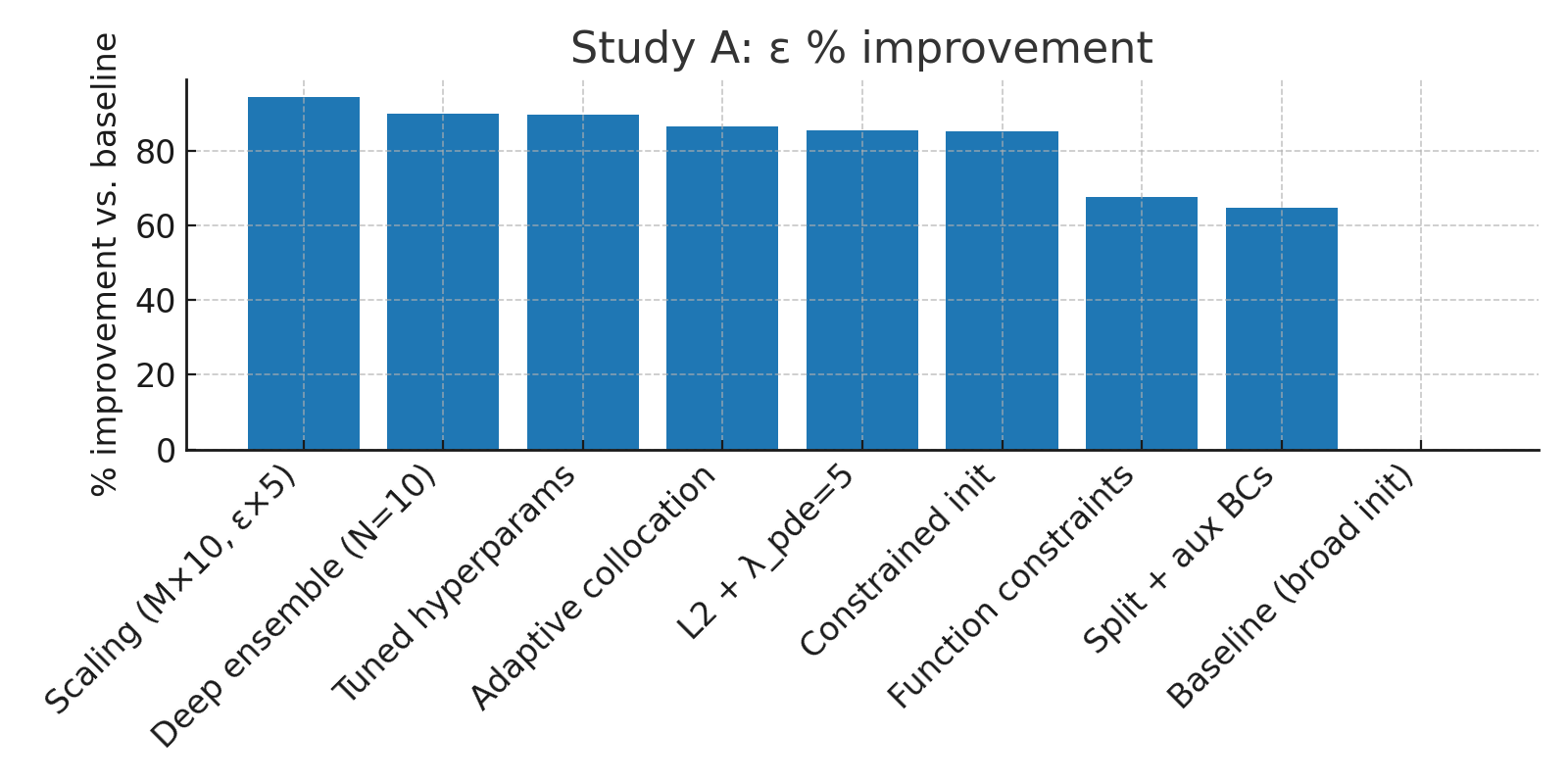}
		\caption{Study A (single trajectory): \% improvement vs.\ the weak baseline for $\mu$, $M$, and $\varepsilon$.}
		\label{fig:studyA_improv}
	\end{figure}

\begin{figure}[p]
		\centering
		\includegraphics[width=0.8\linewidth]{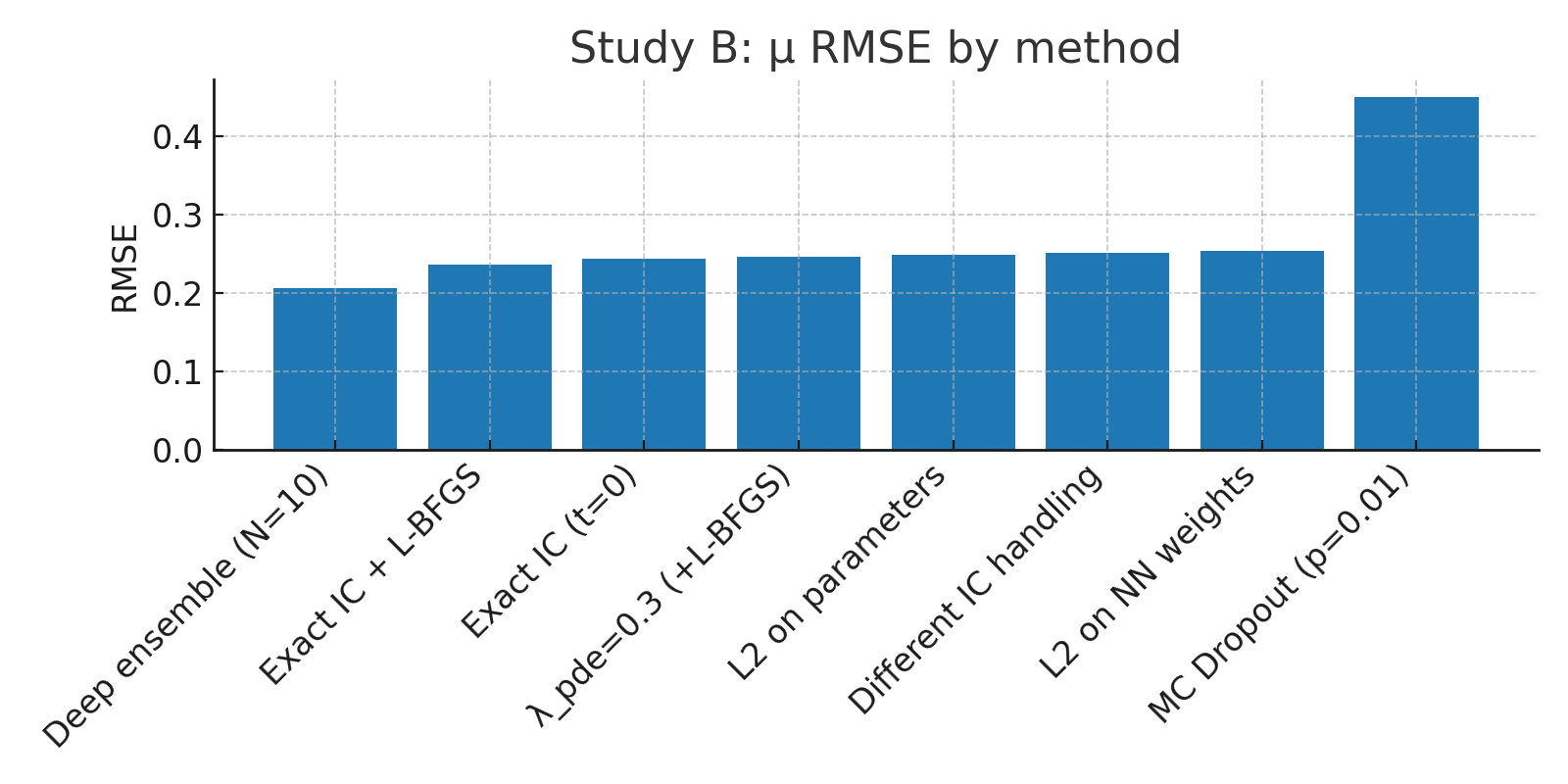}\\
		\includegraphics[width=0.8\linewidth]{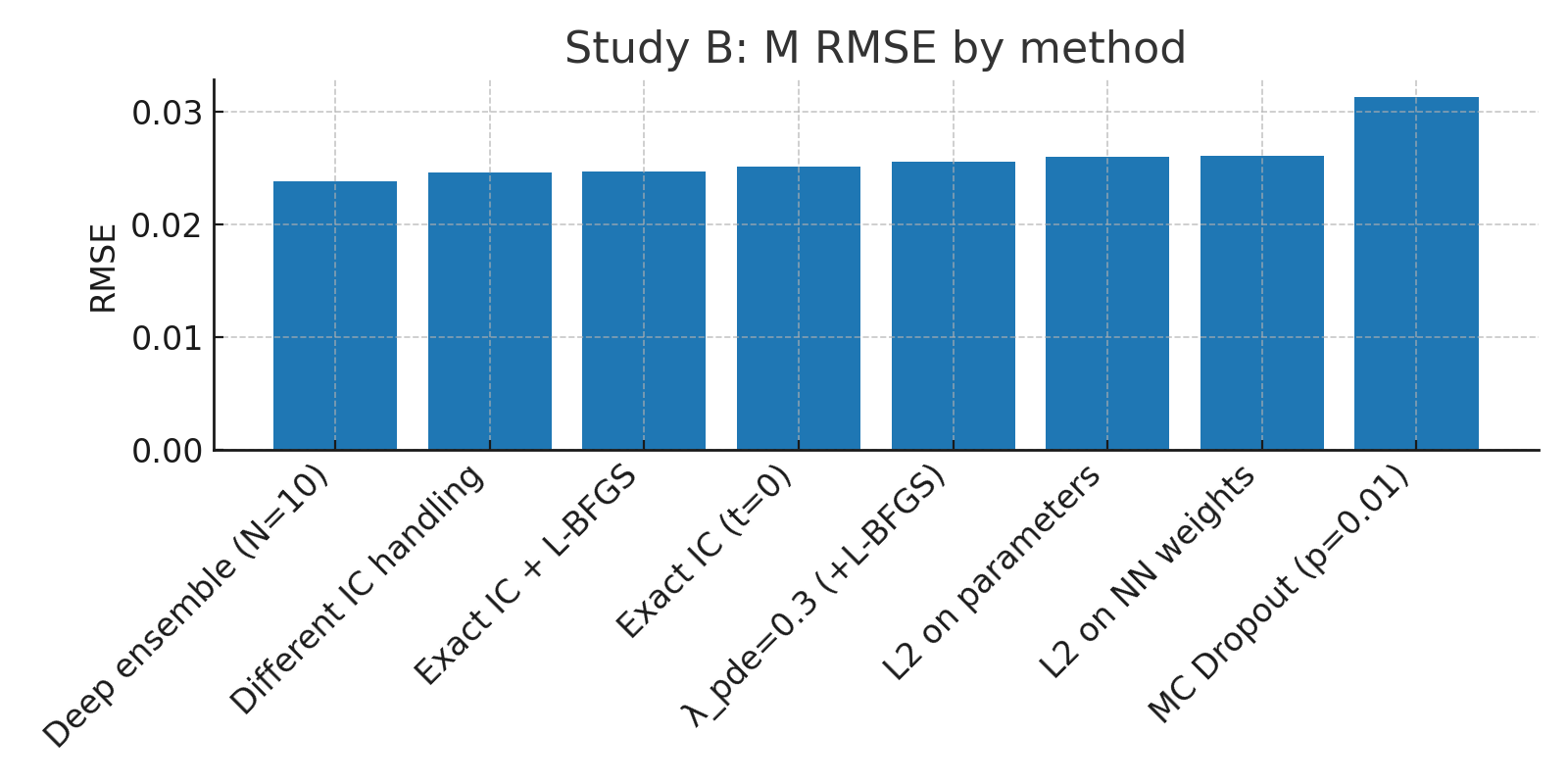}\\
		\includegraphics[width=0.8\linewidth]{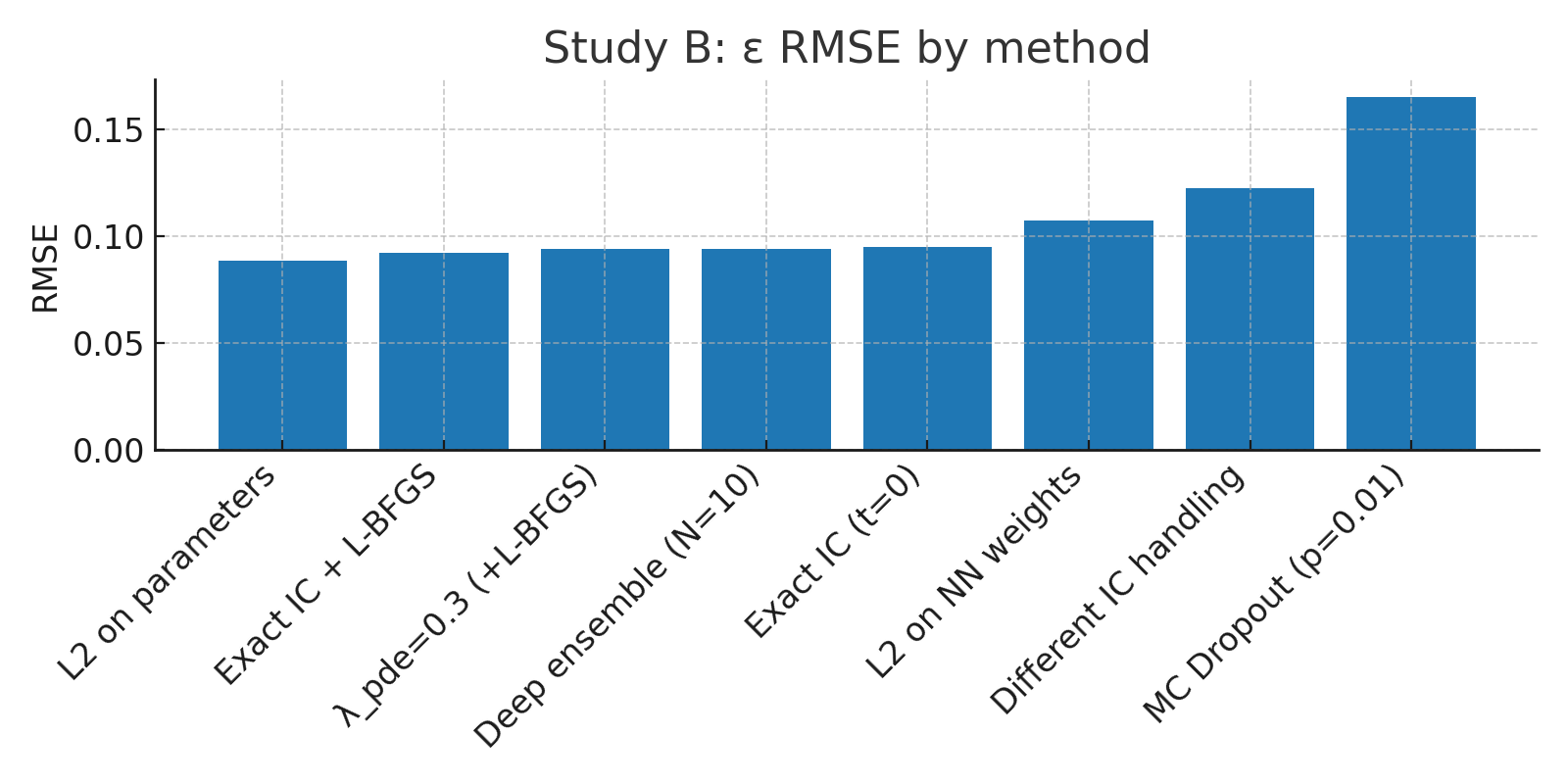}
		\caption{Study B (refined baseline): RMSE by method for $\mu$, $M$, and $\varepsilon$.}
		\label{fig:studyB_rmse}
	\end{figure}

\begin{figure}[p]
		\centering
		\includegraphics[width=0.8\linewidth]{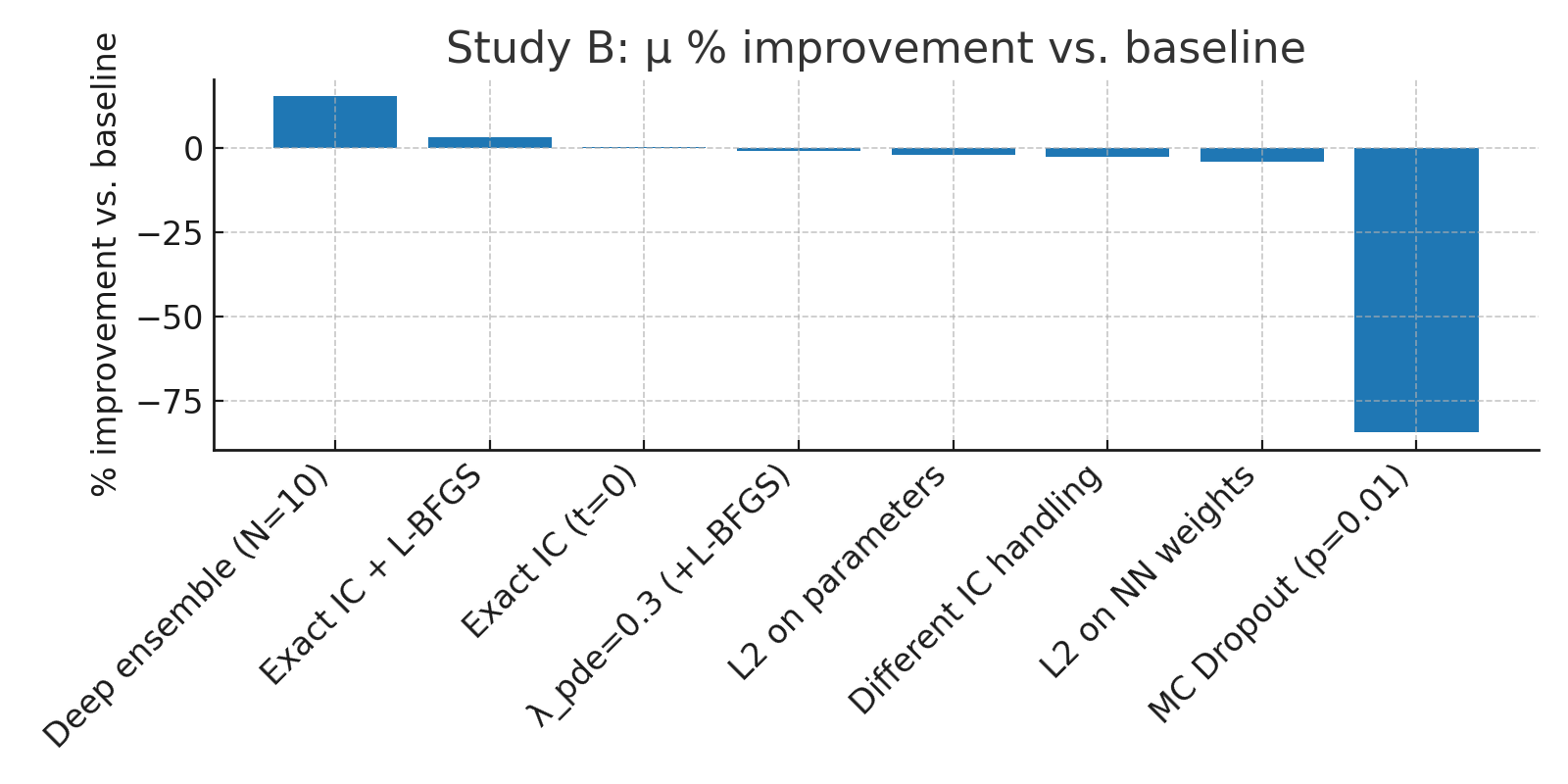}\hfill
		\includegraphics[width=0.8\linewidth]{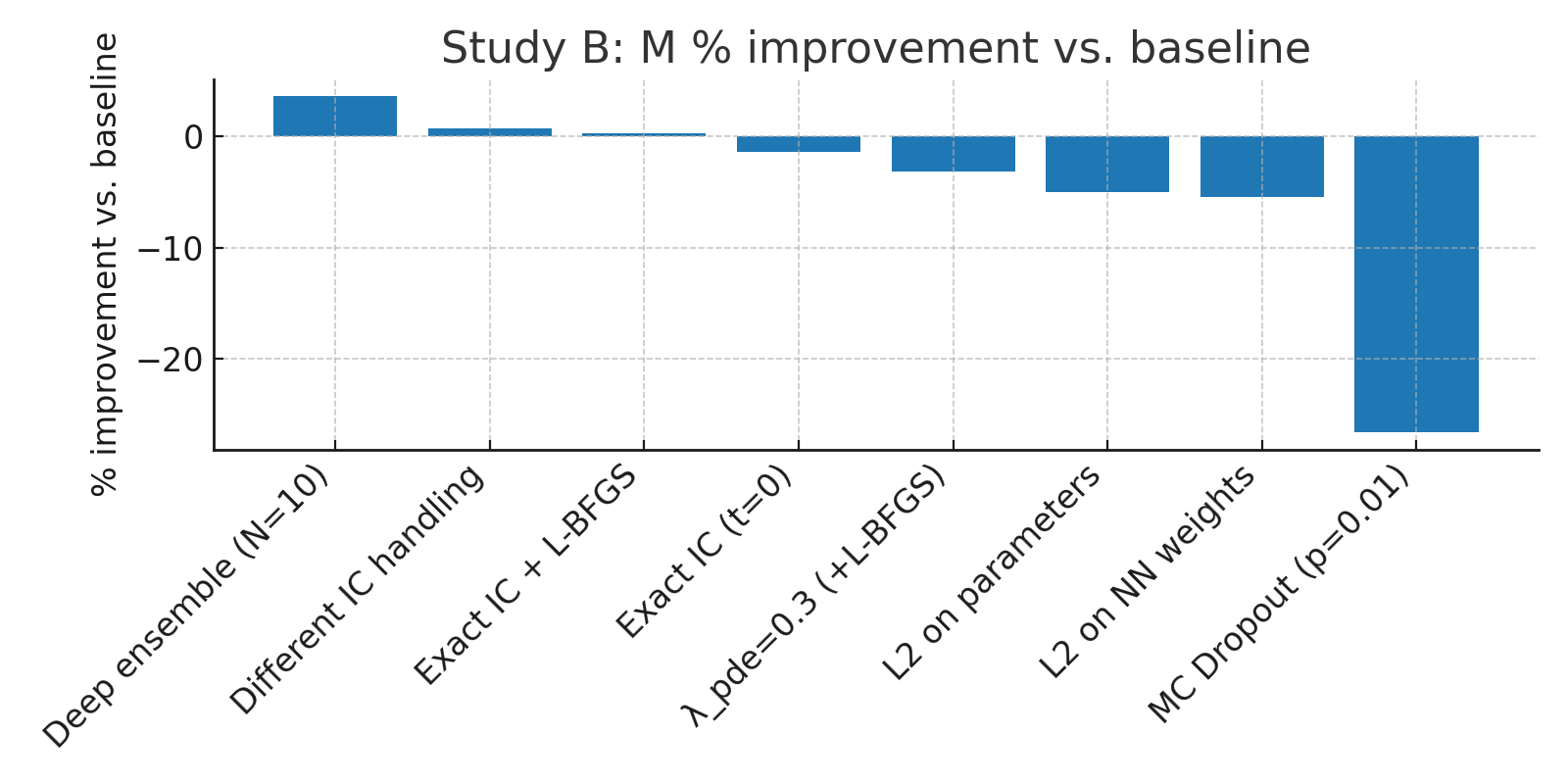}\hfill
		\includegraphics[width=0.8\linewidth]{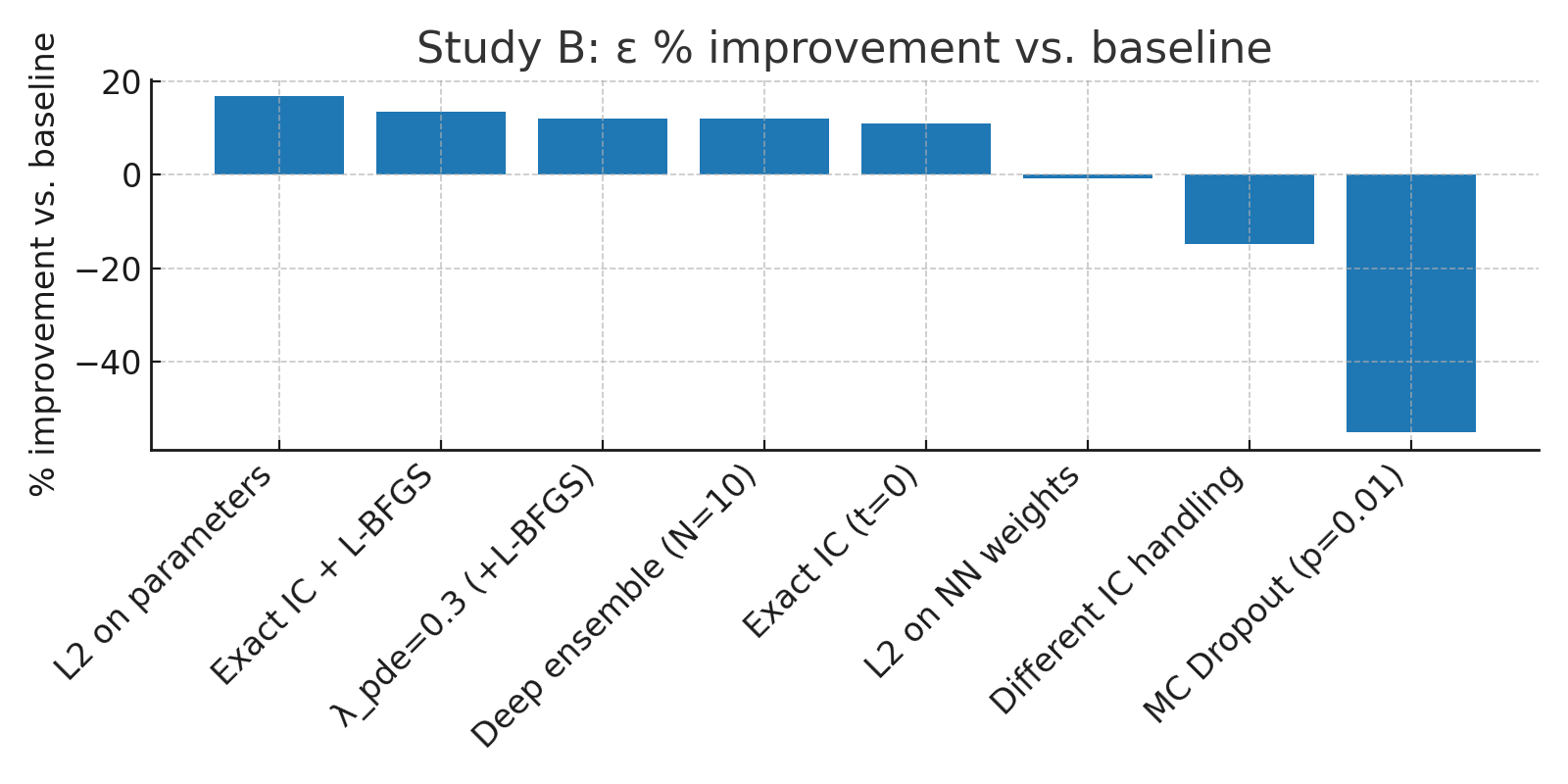}
		\caption{Study B (refined baseline): \% improvement relative to the scaled+FNN baseline.}
		\label{fig:studyB_improv}
	\end{figure}

	We use the ML term “ablation” to mean a controlled variant study in which one design choice (e.g., scaling, loss weight, constraint) is changed at a time while holding all other settings fixed, so observed RMSE differences can be attributed to that choice. Table~\ref{tab:ablation} reports RMSE and percentage improvement vs.\ baseline. Figures~\ref{fig:studyA_rmse}–\ref{fig:studyB_improv} show per-metric RMSE and improvement bar plots.
	
	\begin{table}[t]
		\centering
		\caption{Single-trajectory ablation. Values are RMSE (lower is better); numbers in parentheses are \% improvement vs.\ baseline. Baseline uses broad initialisation, no scaling/constraints.}
		\label{tab:ablation}
		\small
		\begin{tabular}{lccc}
			\toprule
			Method & $\mu$ RMSE & $M$ RMSE & $\varepsilon$ RMSE \\
			\midrule
			Baseline (broad init)                          & 0.405 (0.0\%)  & 1.152 (0.0\%)  & 0.456 (0.0\%) \\
			Constrained parameter inits                    & 0.392 (3.2\%)  & 0.072 (93.8\%) & 0.067 (85.2\%) \\
			Parameter scaling ($M\times10,\ \varepsilon\times5$) & \textbf{0.196} (51.7\%) & \textbf{0.021} (98.2\%) & \textbf{0.025} (94.5\%) \\
			L2 reg + $\lambda_{\text{pde}}{=}5$            & 0.233 (42.5\%) & 0.027 (97.6\%) & 0.067 (85.4\%) \\
			Tuned hyperparameters                          & 0.214 (47.1\%) & 0.025 (97.8\%) & 0.047 (89.8\%) \\
			Adaptive collocation                           & 0.227 (43.9\%) & 0.027 (97.7\%) & 0.061 (86.7\%) \\
			Split training + auxiliary BCs                 & 0.287 (29.1\%) & 0.033 (97.1\%) & 0.161 (64.7\%) \\
			Function constraints ($\mu{>}0$, diag($M$)$<0$)& 0.292 (28.0\%) & 0.034 (97.0\%) & 0.148 (67.6\%) \\
			Deep ensemble ($N{=}10$)                       & \underline{0.185} (54.2\%) & \underline{0.024} (97.9\%) & \underline{0.045} (90.1\%) \\
			\bottomrule
		\end{tabular}
	\end{table}
	
	\subsection{Physics Loss-Weight Sweep}\label{sec:lambda_sweep}
	We varied $\lambda_{\text{pde}}\!\in\!\{0.3,1,3,10\}$ (keeping other settings fixed). Performance was stable for $M$ near $0.025$–$0.028$, while $\mu$ and $\varepsilon$ showed mild sensitivity (Table~\ref{tab:lambda}). 
	
	\begin{table}[t]
		\centering
		\caption{PDE loss-weight $\lambda_{\text{pde}}$ sweep (RMSE).}
		\label{tab:lambda}
		\small
		\begin{tabular}{lccc}
			\toprule
			$\lambda_{\text{pde}}$ & $\mu$ RMSE & $M$ RMSE & $\varepsilon$ RMSE \\
			\midrule
			0.3  & 0.246 & 0.0255 & 0.0937 \\
			1.0  & 0.251 & 0.0264 & 0.1015 \\
			3.0  & 0.259 & 0.0273 & 0.1337 \\
			10.0 & 0.298 & 0.0281 & 0.1261 \\
			\bottomrule
		\end{tabular}
	\end{table}
	
	\begin{figure}[t]
		\centering
		\includegraphics[width=.55\linewidth]{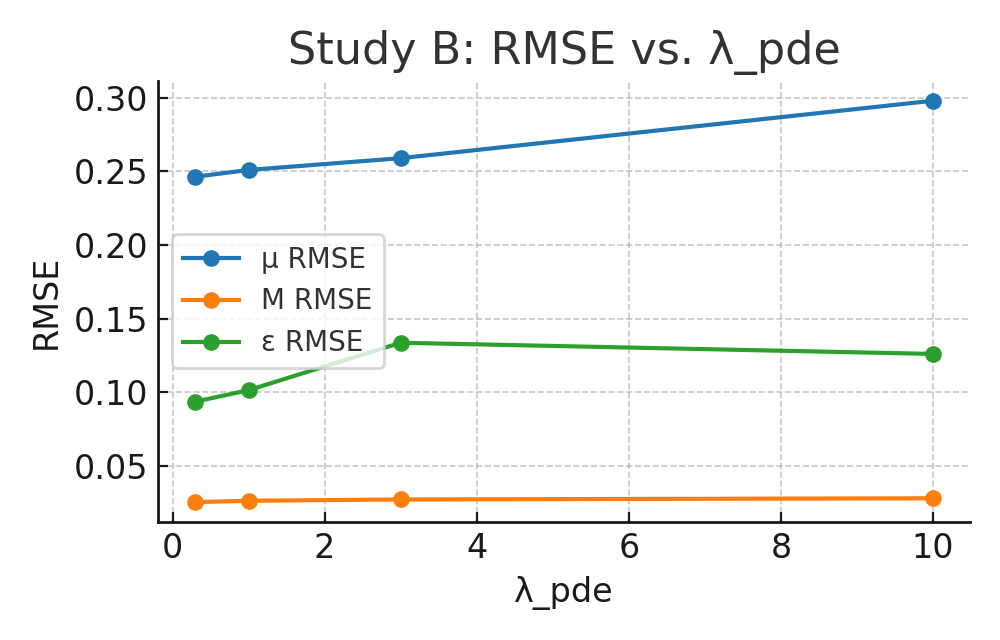}
		\caption{Study B: RMSE as a function of the physics loss weight $\lambda_{\text{pde}}$.}
		\label{fig:lambda_sweep_fig}
	\end{figure}
	
	\subsection{Optimiser \& Uncertainty (additional results)}\label{sec:opt_uq_results}
	Table~\ref{tab:opt-uq} summarises additional settings beyond the core ablation. An Adam$\rightarrow$L-BFGS pass improved $M$ and $\varepsilon$ while keeping $\mu$ competitive. Penalising the parameters themselves (L2 on $\mu,M,\varepsilon$) reduced $\varepsilon$ RMSE. For uncertainty, a deep ensemble remained strong on $\mu$ while MC-dropout bands provided predictive variability at a cost in point estimates here.
	
	\begin{table}[t]
		\centering
		\caption{Additional optimiser/regularisation and UQ results (RMSE).}
		\label{tab:opt-uq}
		\small
		\begin{tabular}{lccc}
			\toprule
			Variant & $\mu$ RMSE & $M$ RMSE & $\varepsilon$ RMSE \\
			\midrule
			Adam $\rightarrow$ L-BFGS & 0.236 & 0.0247 & 0.0922 \\
			L2 on $(\mu,M,\varepsilon)$ & 0.249 & 0.0260 & \textbf{0.0885} \\
			Multi-replicate (different IC; mean over reps) & 0.251 & 0.0246 & 0.122 \\
			Deep ensemble ($N{=}10$; single-trajectory) & \textbf{0.207} & \textbf{0.0238} & 0.0937 \\
			MC-dropout (predictive bands) & 0.451 & 0.0313 & 0.165 \\
			\bottomrule
		\end{tabular}
	\end{table}
	
	\section{Discussion}\label{sec:discussion}
	Two practical takeaways emerge. First, \emph{parameter scaling} is a high-impact, low-effort change that dramatically improves conditioning and accuracy, particularly for $M$ and $\varepsilon$. Second, \emph{deep ensembles} provide the strongest improvements for $\mu$ and help mitigate seed sensitivity at modest extra cost. Adaptive collocation, L2/weighting, and light hyperparameter tuning all contribute incremental gains and are easy to combine with scaling. Loss-weight tuning shows a broad plateau for $M$ in our setup.
	
	Limitations include our emphasis on single-trajectory fits and the well-known structural/practical identifiability challenges of ODE models under noisy, limited data \citep{chis_2011_StructuralIdentifiabilitySystems,villaverde_2016_StructuralIdentifiabilityDynamic,hong_2020_GlobalIdentifiabilityDifferential,wanika_2024_StructuralPracticalIdentifiability}. Function constraints improved feasibility but did not match the best $\mu/\varepsilon$ here; stronger priors or curriculum schedules may help. MC-dropout provides useful predictive bands but often requires calibration or hybridisation with ensembles for reliable uncertainty estimates~\citep{gal2016_dropout,guo2017_calibration,lakshminarayanan2017_deep_ensembles,ovadia2019can}; Methods for total uncertainty quantification and noisy input-output PINN frameworks have been explored and would be interesting to apply in this context ~\citep{zhang2019_total_uncertainty_pinns,zou2023_uq_noisy_pinns}.

	\section{Future Work}\label{sec:future}
	\begin{itemize}
		\item \textbf{Post-hoc / generalized smoothing:} First fit smooth surrogates to each trajectory (e.g., B-splines or GPs), then estimate \ac{gLV} parameters by matching derivatives or penalized residuals (“generalized profiling” / gradient matching). This can (i) decouple denoising from physics, (ii) avoid repeated ODE/PDE solves during early optimization, and (iii) provide fast, stable initialisations for PINN training—often speeding inference while remaining accurate in noisy regimes \citep{ramsay2007parameter,hooker2016collocinfer,dondelinger2013ode}.
		
		\item \textbf{Symbolic regression on residuals:}
		Use PINN residuals $r_i(t)=\dot{x}_i - x_i(\mu_i + \sum_j M_{ij}x_j + \varepsilon_i u(t))$ to discover missing structure (e.g., higher-order, saturating, or context-dependent interactions). Sparse symbolic regression (SINDy / PDE-FIND) on $r_i(t)$ can propose interpretable correction terms that improve fit \emph{and} yield hypotheses about mechanisms \citep{brunton2016_sindy,mangan2017_implicit_sindy,rudy2017_pdefind}.
		
		\item \textbf{Neural ODEs / gray-box hybrids:}
		Benchmark continuous-depth neural dynamics (Neural ODEs) and gray-box “universal differential equations” (known gLV + learned correction $f_\theta(x,t)$) against the PINN on accuracy, speed, and identifiability. Gray-box models preserve mechanistic structure while capturing systematic misspecification, and can amortize inference across conditions \citep{chen2018_neuralode,rackauckas2020_ude,raissi2019physics}.
		
		\item \textbf{Structured priors, sparsity, and curricula:}
		Encourage biologically plausible sparsity in $M$ via L1/group-lasso or Bayesian sparsity priors (spike-and-slab, horseshoe), which improves identifiability in low-signal regimes \citep{tibshirani1996_lasso,mitchell1988_spike_slab,carvalho2010_horseshoe,piironen2017_hs}. Stabilize training with curricula/self-adaptive weighting (gradually ramp $\lambda_{\text{pde}}$ or balance losses) \citep{krishnapriyan2021_pinn_fail,mcclenny2020_sapinn}, and transfer to experimental datasets with appropriate preprocessing (e.g., compositional handling).
		
		\item \textbf{Hybrid UQ:}
		Combine epistemic ensembles with dropout (“ensemble-of-dropouts”) for stronger coverage \citep{gal2016_dropout,lakshminarayanan2017_deep_ensembles}. For calibrated posteriors over parameters, use Bayesian last-layer or Laplace approximations (last-layer or full-network), which are cheap to add to trained models \citep{ritter2018_laplace,daxberger2021_laplace}.
		
	\end{itemize}
	
	\section*{Data and Code Availability}
	All code to reproduce the ablations and the CSV/figures used here will be made publicly available upon submission. A compiled CSV of the RMSE metrics is included as supplementary material (\texttt{studyA\_ablation.csv}, \texttt{studyB\_methods.csv}, \texttt{studyB\_lambda\_sweep.csv}).
	
	\bibliographystyle{plainnat}
	\bibliography{references}
	
\end{document}